\documentclass[namedreferences]{solarphysics}

\usepackage[optionalrh,natbib]{spr-sola-addons}
  \usepackage{graphicx}
  \usepackage{color}
  \usepackage{url}
  \usepackage{a4}
  \usepackage{setspace}

\usepackage{solaheader}	
\newcommand{\solareceiveddate}{9 March 2015} 
\newcommand{\solaaccepteddate}{12 January 2015}

\newcommand{\apj}{    {\it Astrophys. J.}}

\newcommand{\solphys}{{\it Solar Phys.}}

\begin{document}

\begin{article}

\begin{opening}

\title{Estimating electric current densities in solar active regions}

\author{M.~S.~\surname{Wheatland}$^{1}$ 
       }
\runningauthor{Wheatland}
\runningtitle{Estimating current densities in active regions}

\institute{$^{1}$ SIfA, School of Physics, The University of Sydney,
  NSW 2006, Australia, email: \url{m.wheatland@physics.usyd.edu.au}}

\begin{abstract}
Electric currents in solar active regions are thought to provide the 
energy released via magnetic reconnection in solar flares. Vertical 
electric current densities $J_z$ at the photosphere may be estimated 
from vector magnetogram data, subject to substantial uncertainties. 
The values provide boundary conditions for nonlinear force-free 
modelling of active region magnetic fields. A method is presented for 
estimating values of $J_z$ taking into account uncertainties in vector 
magnetogram field values, and minimizing $J_z^2$ across the active 
region. The method is demonstrated using the boundary values of the 
field for a force-free twisted bipole, with the addition of noise at
randomly chosen locations. 
\end{abstract}
\end{opening}

\section{Introduction}
     \label{S-Intro} 

The energy stored in active region magnetic fields produces large-scale
solar activity, in particular solar flares. Flares often occur
at sites in active regions overlying neutral lines where fields are 
strongly sheared, suggesting a connection between flares and large 
scale electric currents in the solar corona.

The most accurate determinations of active region magnetic fields and
associated currents are based on measurements of the 
polarisation state of magnetically sensitive photospheric 
lines~\citep{2003isp..book.....D}. The data are used to 
produce vector magnetograms, which are maps of the vector magnetic 
field ${\bf B}=(B_x,B_y,B_z)$ across regions on the photosphere
(where $x$ and $y$ refer to the heliographic west and north directions
respectively, and $z$ is the radially outwards direction). The process
of going from polarisation measurements to vector magnetogram 
values involves atmospheric modelling, the resolution of an 
intrinsic 180-degree ambiguity in the direction of the component of 
the field transverse to the line of sight, rotation of coordinate 
systems, and often also involves rebinning of the data. The resulting
field values are substantially uncertain. Contributions to the 
uncertainty include errors in the original
polarisation measurements, inaccuracy of the atmospheric model 
assumed in spectro-polarimetric inversion, errors in the ambiguity 
resolution, and loss of information in rebinning. 

The measurement errors depend on the instrument, and two types of 
instruments are in use. Spectro-polarimeters record the full 
Stokes profile at many wavelengths across a magnetically sensitive 
spectral line, whilst magnetographs, or filter-based instruments, 
record polarisation signals integrated with respect to wavelength over 
a line \citep{LanLan04}. The Hinode Solar Optical Telescope (SOT) Focal 
Plane Package includes a Spectro-Polarimeter~\citep{TsunetaEtAl2008}, 
and the Helioseismic and Magnetic Imager on the Solar Dynamic Observatory 
(SDO) provides filtergrams~\citep{2012SoPh..275..207S}.
Spectro-polarimeter typically have a high spectral resolution and 
provide very accurate measurements of the polarisation state, but they
take longer to produce the data for a vector magnetogram and the
magnetic field may change during the observation interval. Filter 
instruments produce data for a region on the Sun more rapidly, but 
they lack the spectral resolution and intrinsic high accuracy provided 
by a spectro-polarimeter.

Given a set of vector magnetic field values, estimates of the vertical
(locally radial) electric current density at the level of the 
photosphere ($z=0$) may be made using Amp\'{e}re's law:
\begin{equation}\label{eq:jz0_ampere}
\mu_0 J_z (x,y,0) =v-u,
\end{equation}
where we introduce the notation
\begin{equation}\label{eq:defn_uv}
u=\left.\frac{\partial B_x}{\partial y}\right|_{z=0}
\quad \mbox{and}\quad
v=\left.\frac{\partial B_y}{\partial x}\right|_{z=0}.
\end{equation}
If the uncertainties in the field values are $\sigma_x$ and $\sigma_y$,
and centered differencing is used to estimate the derivatives, then
the uncertainties in the derivatives are $\sigma_u=\sigma_x/\sqrt{2}h$
and $\sigma_v=\sigma_y/\sqrt{2}h$, where $h$ is the grid spacing in 
the magnetogram in $x$ and $y$. If the spatial scale for variation of 
the horizontal field is $L$, then we have $u\sim \overline{B}_x/L$ and
$v\sim \overline{B}_y/L$, where $\overline{B}_x$ and $\overline{B}_y$
denote characteristic values of the field components. In that case
\begin{equation}\label{eq:unc_deriv_ests}
\frac{\sigma_u}{u}\sim \frac{L}{h}\frac{\sigma_x}{\overline{B}_x}
\quad \mbox{and}\quad
\frac{\sigma_v}{v}\sim \frac{L}{h}\frac{\sigma_y}{\overline{B}_y}.
\end{equation}
Equations~(\ref{eq:unc_deriv_ests}) imply that if there are 
significant fractional uncertainties in the field values, then the 
derivative estimates may be substantially uncertain, in
particular for high-resolution data (with $L\gg h$). The current 
densities obtained using Equation~(\ref{eq:jz0_ampere}) will be
correspondingly uncertain.

Errors in the determination of current densities from vector 
magnetogram data were discussed by \cite{1994ApJ...425L.117P} and 
\cite{1999SoPh..188....3L}, who assigned constant uncertainties 
for field components in the directions along, and transverse to, 
the line of sight. Uncertainty values may be
assigned to individual vector magnetogram field values based on the 
quality of the model fit in the spectro-polarimetric inversion, 
and this information is now routinely supplied with vector magnetogram 
field values (e.g.\ \citealp{2014SoPh..289.3483H}).
Other sources of uncertainty have also been discussed, including
errors in ambiguity resolution~\citep{2009SoPh..260...83L},
and errors introduced by the rebinning of 
data in the construction of magnetograms~\citep{2012SoPh..277...89L}.

The nonlinear force-free model is often used to model the coronal 
magnetic field based on vector magnetogram 
data (e.g.\ \citealp{2012LRSP....9....5W}). The model is defined by
${\rm curl}\,{\bf B}=\alpha {\bf B}$ and ${\rm div}\,{\bf B} = 0$,
where $\alpha$ is the force-free parameter. The \citet{gr1958} 
method of solution of these equations takes as boundary conditions the
values of $B_z$ at the photosphere, and the values of $J_z$ (or the 
values of $\alpha$) over one polarity of $B_z$. Other methods, in 
particular the optimization 
method (e.g.\ \citealp{2000ApJ...540.1150W}, 
\citealp{2010A&A...516A.107W}), use all
three components of ${\bf B}$ at the photosphere as boundary 
conditions. In practice all methods modify the
boundary data substantially from the values in the vector
magnetogram in order to achieve a solution to the force-free 
model~(see e.g.\ \citealp{2009ApJ...696.1780D}, 
\citealp{2011ApJ...728..112W}, \citealp{2012SoPh..281...37W}). 
For the active region 
modelled by \citet{2011ApJ...728..112W} using the `self-consistent' 
Grad-Rubin procedure \citep{2009ApJ...700L..88W}, the current 
densities in the magnetogram 
were altered to the extent that the average absolute change in the 
horizontal field was 170~gauss, and the average ratio of the change
in the horizontal field to the assigned uncertainty in the horizontal 
field, on a pointwise basis across the magnetogram, was $\approx 9$. 
The changes greatly exceeded the nominal uncertainties, which is
typical. The `preprocessing' procedure often used with the optimization
method of nonlinear force-free 
solution~\citep{2006SoPh..233..215W}
alters the transverse components of the field by
$\lesssim 500$~gauss in strong field regions~\citep{2011A&A...526A..70F}. 
If nonlinear force-free codes are applied to vector magnetogram 
boundary data without these changes, accurate solutions to the 
nonlinear force-free model are (generally) not obtained. 

The large changes required in vector magnetogram data to achieve 
nonlinear force-free model solutions are most likely due to the 
inconsistency of boundary data with the model~(see e.g.\ 
\citealp{2008ApJ...675.1637S}, \citealp{2009ApJ...696.1780D}). The
solar magnetic field is not expected to be force-free at the level
of the photosphere. Errors in the field estimates also lead to 
inconsistency. Depending on the solution method, this problem may cause
the solution magnetic fields to have significant departures from the
divergence-free condition~\citep{2013A&A...553A..38V}.

In this paper we reconsider the problem of estimating electric current
densities from vector magnetogram boundary data, taking into account
uncertainties in the data. A method is introduced for calculating a
set of boundary values of $J_z$ which minimize departures from centred 
difference estimates for the current densities at locations where the 
values are accurate (according to the uncertainties), and which 
minimize the sum of $J_z^2$ across the magnetogram. The goal is an
estimate of the current density which avoids large values produced by
errors. The method is tested in application to a test case with known
errors. The paper is divided as follows. Section~2 presents the method,
and Section~3 describes the test and analyzes the results. Section~4 
draws conclusions.

\section{Method}
     \label{S-Method} 

Equation~(\ref{eq:jz0_ampere}) provides the vertical current density
in terms of the horizontal field gradients $u$ and $v$. The problem 
consists in estimating the gradients from the data, subject to 
unertainties, and we also seek to avoid large (spurious) values of 
the resulting vertical current densities, produced by points with
large uncertainties. Hence we consider the problem of minimizing
\begin{equation}\label{eq:F}
F=\sum_{ij}\frac{(u_{ij}-u_{ij}^{\rm est})^2}{2\sigma_{u\,ij}^2}+
  \frac{(v_{ij}-v_{ij}^{\rm est})^2}{2\sigma_{v\,ij}^2}+
  \lambda \sum_{ij}J_{z\,ij}^2,
\end{equation}
where the indices refer to points $(x_i,y_j)$ in the magnetogram,
$u_{ij}^{\rm est}$ and $v_{ij}^{\rm est}$ are estimates of 
the gradients, and where $\lambda$ is a constant.
It is useful to non-dimensionalize by expressing lengths, field 
strengths, and current densities in units of characteristic 
values $B_s$, $L_s$, and $B_s/\mu_0L_s$, respectively. Using 
Equation~(\ref{eq:jz0_ampere}), Equation~(\ref{eq:F}) may be written
in non-dimensional form
\begin{equation}\label{eq:F_nondim}
F=\sum_{ij}\frac{(u_{ij}-u_{ij}^{\rm est})^2}{2\sigma_{u\,ij}^2}+
  \frac{(v_{ij}-v_{ij}^{\rm est})^2}{2\sigma_{v\,ij}^2}+
  \Lambda \sum_{ij}\left(v_{ij}-u_{ij}\right)^2,
\end{equation}
where $\Lambda=\lambda B_s^2/\mu_0L_s^2$.

A minimum value of $F$ is obtained when 
$\partial F/\partial u_{ij}=\partial F/\partial v_{ij}=0$, which
gives two coupled linear equations for the values of $u_{ij}$ and
$v_{ij}$ corresponding to the minimum:
\begin{eqnarray}\label{eq:uvmin}
\frac{u_{ij}^{\rm min}-u_{ij}^{\rm est}}{\sigma_{u\,ij}^2}
  -2\Lambda\left(v_{ij}^{\rm min}-u_{ij}^{\rm min}\right) &= 0 \nonumber \\
\frac{v_{ij}^{\rm min}-v_{ij}^{\rm est}}{\sigma_{v\,ij}^2}
  +2\Lambda\left(v_{ij}^{\rm min}-u_{ij}^{\rm min}\right) &= 0.
\end{eqnarray}
The simultaneous solution of Equations~(\ref{eq:uvmin}) is given by
\begin{equation}\label{eq:umin}
u_{ij}^{\rm min} =\frac{1}{d}\left[
  \left(\frac{1}{\sigma_{v\,ij}^2}+2\Lambda\right)
  \frac{u_{ij}^{\rm est}}{\sigma_{u\,ij}^2}
  +2\Lambda \frac{v_{ij}^{\rm est}}{\sigma_{v\,ij}^2}\right] 
\end{equation}
and
\begin{equation}\label{eq:vmin}
v_{ij}^{\rm min} =\frac{1}{d}\left[
  2\Lambda \frac{u_{ij}^{\rm est}}{\sigma_{u\,ij}^2}
  +\left(\frac{1}{\sigma_{u\,ij}^2}+2\Lambda\right)
  \frac{v_{ij}^{\rm est}}{\sigma_{v\,ij}^2}\right],
\end{equation}
where
\begin{equation}\label{eq:det}
d=\frac{1}{\sigma_{u\,ij}^2\sigma_{v\,ij}^2}+2\Lambda
  \left(\frac{1}{\sigma_{u\,ij}^2}+\frac{1}{\sigma_{v\,ij}^2}\right).
\end{equation}
The current density corresponding to the gradients which minimize $F$ is 
$J_{z\,ij}^{\rm min}=v_{ij}^{\rm min}-u_{ij}^{\rm min}$, which 
evaluates to
\begin{equation}\label{eq:jzmin}
J_{z\,ij}^{\rm min}=\frac{J_{z\,ij}^{\rm est}}{1+2\Lambda\sigma_{J\,ij}^2},
\end{equation}
where 
\begin{equation}\label{eq:def_jzd}
J_{z\,ij}^{\rm est}=v_{ij}^{\rm est}-u_{ij}^{\rm est}
\end{equation}
and
\begin{equation}\label{eq:def_sigjz}
\sigma_{J\,ij}^2=\sigma_{u\,ij}^2+\sigma_{v\,ij}^2. 
\end{equation}
Equation~(\ref{eq:jzmin}) provides a surprisingly simple 
solution to the problem: the current density at each point is 
reduced from the usual estimate by a factor 
\begin{equation}
f_{ij}=\left(1+2\Lambda\sigma_{J\,ij}^2\right)^{-1},
\end{equation}
so at points with larger uncertainties the current density is reduced 
more. We refer to Equation~(\ref{eq:jzmin}) as the `minimum current'
estimate for the current density.

In the following we use centered differences to provide the estimates
of the gradients:
\begin{equation}\label{eq:uijd}
u_{ij}^{\rm est}=\frac{B_x(x_i,y_j+h)-B_x(x_i,y_j-h)}{2h}
\end{equation}
and
\begin{equation}\label{eq:vijd}
v_{ij}^{\rm est}=\frac{B_y(x_i+h,y_j)-B_y(x_i-h,y_j)}{2h},
\end{equation}
where $h$ is the grid spacing in the magnetogram in $x$ and $y$.
With these choices the uncertainties in the gradients, using 
propagation of errors, are
\begin{equation}\label{eq:su}
\sigma_{u\,ij}=\frac{1}{2h}
  \left(\sigma_{x\,ij+1}^2+\sigma_{y\,ij-1}^2\right)^{1/2},
\end{equation}
and 
\begin{equation}\label{eq:sv}
\sigma_{v\,ij}=\frac{1}{2h}
  \left(\sigma_{y\,i+1j}^2+\sigma_{y\,i-1j}^2\right)^{1/2},
\end{equation}
and the error in the current density is
\begin{equation}\label{eq:sjz}
\sigma_{J\,ij}=\frac{1}{\sqrt{2}h}
  \left(\sigma_{x\,ij+1}^2+\sigma_{y\,ij-1}^2+\sigma_{y\,i+1j}^2
  +\sigma_{y\,i-1j}^2\right)^{1/2}.
\end{equation}
Higher order differencing schemes are sometimes used. In the general 
case Equation~(\ref{eq:uijd}) is replaced by 
\begin{equation}\label{eq:uijd-general}
u_{ij}^{\rm est}=\frac{1}{2h}\sum_k c_kB_x(x_i,y_j+kh),
\end{equation}
where the sum enumerates the points involved in the differencing, and 
the $c_k$ are coefficients. Equation~(\ref{eq:su}) is replaced by
\begin{equation}\label{eq:su-general}
\sigma_{u\,ij}=\frac{1}{2h}\sum_k
  \left(\alpha_k^2\sigma_{x\,ij+k}^2\right)^{1/2},
\end{equation}
and corresponding relations hold for $v_{ij}^{\rm est}$ and 
$\sigma_{v\,ij}$. Equation~(\ref{eq:sjz}) is replaced by
\begin{equation}\label{eq:sjz-general}
\sigma_{J\,ij}=\frac{1}{2h}
  \left[\sum_k\alpha_k^2
  \left(\sigma_{x\,ij+k}^2+\sigma_{y\,ij+k}^2\right)
  \right]^{1/2}.
\end{equation}

\section{Test using simulated data}
     \label{S-Test} 

{\bf 
\subsection{Boundary conditions and minimum current estimate}
}

The method is demonstrated in application to a test case consisting 
of boundary conditions for a twisted bipole, with noise added to 
the boundary values at some points.

Panels (a), (b), and (c) of Figure 1 illustrate the twisted-bipole
boundary conditions. Panel (a) shows the vertical component of
the boundary field, $B_z(x,y,0)$, which is constructed by adding the 
field from monopole sources at locations $(x_{+},y_{+},z_{+})$ and 
$(x_{-},y_{-},z_{-})$. In non-dimensional units we have
\begin{equation}\label{eq:bz0_bcs}
B_z(x,y,0) = C\left(
  \frac{z-z_{+}}{R_{+}^3}-\frac{z-z_{-}}{R_{-}^3}
  \right),
\end{equation}
with
\begin{equation}
R_{\pm}=\left[(x-x_{\pm})^2+(y-y_{\pm})^2+z_{\pm}^2\right]^{1/2},
\end{equation}
and with $C$ chosen such that ${\rm max}\left|B_z(x,y,0)\right|=1$. 
The boundary field is constructed over the region $0\leq x \leq 1$ and 
$0\leq y\leq 1$, with the choices $x_{+}=0.6$, $x_{-}=0.4$, 
$y_{+}=y_{-}=0.5$, and $z_{+}=z_{-}=-0.2$. A current is introduced
by assuming a boundary distribution of the force-free parameter
$\alpha$ over the positive polarity of the field. We choose
\begin{equation}\label{eq:bcJ}\label{eq:alpha0_bcs}
\alpha (x,y,0) =\left\{\begin{array}{ll}
    \alpha_0 & \mbox{if $B_z\geq B_1$} \\
    0 & \mbox{otherwise},
    \end{array}
    \right.
\end{equation}
with $B_1=0.9$ and $\alpha_0=15$. The boundary conditions on $B_z$
and $\alpha$ are used to calculate a nonlinear force-free field
in the cubical region defined by the boundary region and 
$0\leq z\leq 1$, using the CFIT Grad-Rubin 
code~\citep{2007SoPh..245..251W}.
The calculation is performed on a $100\times 100\times 100$ grid. 
Panel (b) of Figure~1 shows the magnitude of the horizontal field
\begin{equation}
B_h(x,y,0)=\left[B_x(x,y,0)^2+B_y(x,y,0)^2\right]^{1/2}
\end{equation}
for the calculated force-free field and panel (c) shows the 
corresponding values of the current density $J_z (x,y,0)$ for the 
force-free field, estimated using Equation~(\ref{eq:jz0_ampere}) 
with the centered difference values
Equations~(\ref{eq:uijd}) and~(\ref{eq:vijd}) used for the field 
gradients. The patch of positive current matches the current density
$J_z =\alpha B_z$ defined by the boundary conditions on $B_z$ and 
$\alpha$ assumed for the force-free calculation 
[Equations~(\ref{eq:bz0_bcs}) and~(\ref{eq:alpha0_bcs})], and the 
patch of negative current density follows from the mapping of the 
boundary values of $\alpha$ along field lines in the force-free field.

Noise is added to the boundary values of ${\bf B}$ from the force-free
calculation, to provide a set of `observed' boundary values. At a 
fraction $\theta$ of the boundary positions, randomly chosen, 
additional components $\delta B_x$ and $\delta B_y$ are added to 
$B_x(x,y,0)$ and $B_y(x,y,0)$ respectively, where $\delta B_x$ and 
$\delta B_y$ are normally distributed random numbers with mean zero
and standard deviation $\sigma_x=\sigma_y=\sigma_0$, and where
$\sigma_0$ is a chosen constant. The boundary values of $B_z (x,y,0)$ 
are not altered. Panel (d) of Figure~1 illustrates the resulting
noisy values of the horizontal field, for the choices $\theta=0.1$ 
and $\sigma_0=0.25$, and panel (e) shows the corresponding current
density, which we use for $J_z^{\rm est}$. The values are obtained
using Equation~(\ref{eq:jz0_ampere}) applied to the observed (noisy)
boundary values with centered differencing used to estimate the 
gradients. The noise in the horizontal field leads to spurious values 
of the current density, which mask the positive and negative patches 
seen in panel (c).

Panel (f) of Figure~1 shows the minimum current estimate 
$J_z^{\rm min}$ for the current density calculated using 
Equation~(\ref{eq:jzmin}) with the choice $\Lambda = 5$. 
This value of $\Lambda$ was chosen because it reduces 
the noise in weak field regions to small values. The values 
of $J_z(x,y,0)$ shown in panel (e) provide the estimate 
$J_z^{\rm est}$, and the uncertainties in the current density values 
$\sigma_J$ follow from Equations~(\ref{eq:su})-(\ref{eq:sv}) assuming 
$\sigma_x=\sigma_y=\sigma_0$ at points with noise, and 
$\sigma_x=\sigma_y=0$ at points without noise. Panel (f) shows 
that $J_z^{\rm min}$ returns a good approximation to the underlying 
distribution of current density in the boundary, except at locations 
affected by the noise in the horizontal field. 

With the chosen value of $\Lambda$, the RMS value of
$(J_z^{\rm est}-J_z^{\rm min})/\sigma_J$ 
over the boundary points with $\sigma_J>0$ is close to unity. This
is because the current density at the locations affected by noise is 
reduced by a value essentially equal to the noise. For this test case 
the minimum current estimate changes the boundary values within the
uncertainties.

{\bf 
\subsection{NLFFF reconstructions}
}

The minimum current estimates for the current density enable
reconstruction of a nonlinear force-free field.
If the CFIT code is applied using the noisy current density estimates
$J_z^{\rm est}$ the Grad-Rubin iteration does not converge: the noise
in the data prevents a successful reconstruction. However, accurate
force-free solutions are obtained using the minimum current estimate
for the current densities.

Figure~2 illustrates the force-free solutions. Panel (a)
shows the original NLFFF obtained from the boundary data without 
noise. The greyscale shows the boundary values of $B_z$, which are 
shown also in panels (b) and (d), and the white curves are a set of 
field lines originating at the positive pole. (Each of the panels
in Figure~2 shows a cropped central region, by comparison with 
Figure~1.) Panel (b) shows the P and N solutions obtained from the 
boundary values of $B_z$ together with the minimum current method 
estimates for current density. The white field lines correspond to 
the P solution, and the black field lines to the N solution. Panel 
(b) shows that the P and N solutions are twisted bipole configurations
which are similar to the original field [panel (a)] and to each 
other, but are not identical. The P and the N solutions have 
less twist than the original field due to the reductions in current 
density at points affected by noise, and the energies of the fields 
are correspondingly decreased. The energy $E$ of the original 
field shown in panel (a) is given by $E/E_0=1.083$, where $E_0$ 
is the energy of a potential field with the same lower boundary 
values for $B_z$ [calculated by the method described for the CFIT 
code~\citep{2007SoPh..245..251W}]. The energies of the P and N 
solutions shown in panel (b) are given by $E_{\mathrm{P}}/E_0=1.064$
and $E_{\mathrm{N}}/E_0=1.048$. The reconstructed P and N fields 
contain about 80\% and about 60\% of the free magnetic energy of
the original field, respectively.

We apply also the self-consistency procedure presented
in \cite{2009ApJ...700L..88W}, which uses boundary values at
both polarities to identify a single NLFFF solution. The procedure 
involves a cycle of calculations of P and N solutions. After a pair
of solutions is calculated, the boundary values of $\alpha$ in the two
solutions are averaged, taking into account uncertainties in the
$\alpha$ values. The resulting averaged $\alpha$ values provide the 
boundary conditions for the P and N solutions in the next cycle.
After a number of cycles, the P and N solutions agree, and provide
a `self consistent' solution. Panels (c) and (d) in Figure 2 
illustrate the results. Panel (c) shows the P and N solutions -- the
white and black curves, respectively -- after nine self-consistency 
cycles. The two sets of field lines are mostly overlapping. The
energies of the P and N solutions are 
$E_{\mathrm{P}}/E_0=E_{\mathrm{N}}/E_0=1.085$. A self-consistent solution
is found, with a free energy a few percent higher than the 
energy of the original field. Panel (d) shows the $J_z$ values in 
the self-consistent field, for comparison with panel (c) of Figure~1.
The self-consistency procedure recovers a good approximation to
the original set of current density values, although there are 
detailed differences. The method is able, to some extent, to `remove 
the noise' because at a boundary point affected by noise, say at 
the P polarity, the uncertainty in the $\alpha$ value is large. The 
averaging procedure replaces this $\alpha$ value with the $\alpha$ 
value from the N solution, provided it has a much smaller 
uncertainty. This procedure generally recovers the original 
$\alpha$ values.

The self-consistent solution is qualitatively similar 
to the original field. A quantitative measure is provided by the 
mean vector error \citep{2006SoPh..235..161S}:
\begin{equation}
E_m =\frac{1}{M}\sum_i
\frac{\left|{\bf B}_{\mathrm{orig},i}-{\bf B}_{\mathrm{sc},i}\right|}
  {\left|{\bf B}_{\mathrm{sc},i}\right|},
\end{equation}
where ${\bf B}_{\mathrm{orig}}$ is the original NLFFF,
${\bf B}_{\mathrm{sc}}$ is the field obtained by the self-consistency
procedure, and $M=100^3$ is the number of grid points. We find 
$E_m=1.6\times 10^{-2}$, indicating close correspondence.

\begin{figure}
\centerline{\includegraphics[width=\textwidth,clip=]{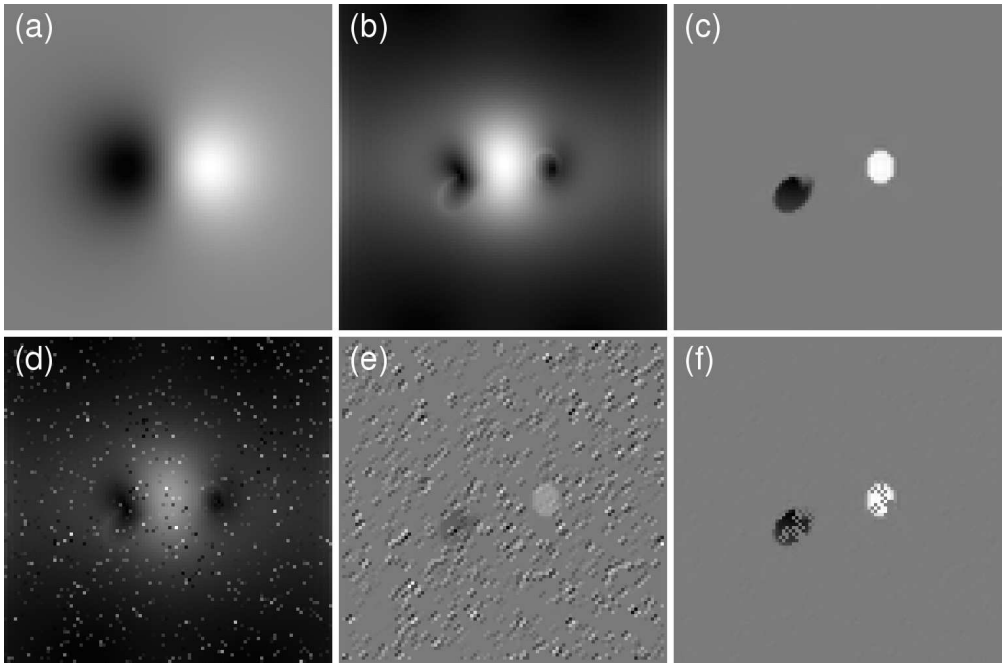}}
\caption{Application of the method to a simple bipole test case: 
boundary values. Panel (a) shows the boundary conditions on $B_z$ for
the test case and panel (b) shows the 
horizontal field in the boundary, obtained from a NLFFF solution
with the chosen boundary conditions on $B_z$ and current density.
Panel (c) shows values of $J_z$ in the boundary for the force-free 
field. Noise is introduced to the horizontal field values at a fraction
of boundary locations, as shown in panel (d), which influences the 
estimated values for $J_z$ [panel (e)]. Panel (f) is the minimum 
current estimate $J_z^{\rm est}$ for the vertical current density
from the noisy boundary data. }
\label{fig:f1}
\end{figure}

\begin{figure}  
\centerline{\includegraphics[width=0.9\textwidth,clip=]{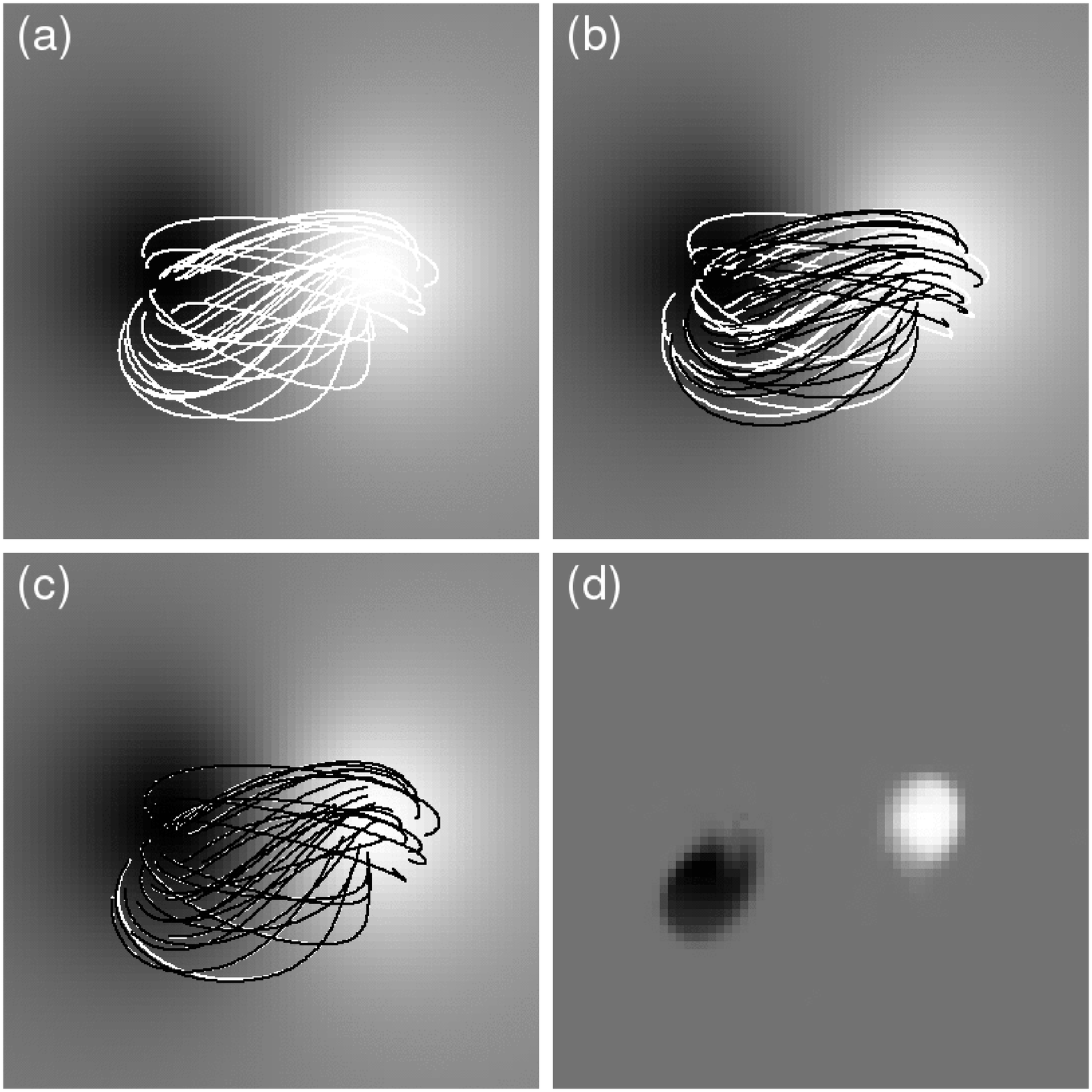}}
\caption{Application of the method to a simple bipole test case: 
nonlinear force-free fields. Panel (a) shows a set of field 
lines for the force-free field calculated from the original bipole 
boundary data, originating at the positive pole.
Panel (b) shows comparable sets of field lines for the P solution 
(white lines) and for the N solution (black lines), calculated from the
bipole values of $B_z$ together with the minimum current method 
estimates $J_z^{\rm min}$ for the current density over the 
P and N polarities of the field in the boundary, respectively.
Panel (c) shows the solutions obtained with the 
Wheatland and Regnier (2009) self-consistency procedure 
(the white/black field lines show the P/N solutions at the end 
of a set of self-consistency cycles). Panel (d) shows the boundary
values for current density in the self-consistency solution. 
The images in each panel are cropped to a central region
in the computational domain, to more clearly show the field 
lines.
\label{fig:f2}}
\end{figure}

\vspace{0.3cm}

\section{Discussion and conclusions}

A method is presented (the `minimum current' method) for estimating 
vertical electric current densities {\bf $J_z$} at the photosphere 
from vector magnetogram data.
The procedure alters the horizontal field gradients used to calculate 
the current densities subject to the uncertainties in the estimates for 
the field gradients, and also minimizes the sum of the square of the 
vertical current densities over all points in the magnetogram. The 
resulting minimum current estimate for the vertical current density 
at a given boundary location is reduced from the original estimate 
by a factor 
$1+\Lambda \sigma_J^2$, where $\sigma_J$ is the uncertainty in the 
original estimate of the current density,
and $\Lambda$ is a chosen constant which determines the relative 
importance of adherence to the original current density estimate or 
minimization of the sum of $J_z^2$ over all boundary locations.

The minimum current method provides an estimate for the current 
density which preserves values of the current density at locations with 
small uncertainties, and reduces the current density at locations with
large uncertainties. The nonlinear force-free (NLFFF) model is 
often applied to vector magnetogram data to provide a proxy coronal 
field~(e.g.\ \citealp{2012LRSP....9....5W}). The construction of an 
accurate solution to the model requires modification of the boundary 
data from observed magnetogram values~(e.g.\ 
\citealp{2008ApJ...675.1637S}, \citealp{2009ApJ...696.1780D}). The 
minimum current method may provide a more systematic approach to this 
modification.

A demonstration of the method is presented, in application to 
boundary field values for a NLFFF twisted bipole field 
configuration, with noise of a 
known amplitude added at a fraction of boundary locations. The model 
field values are taken to represent observed data with known 
uncertainties. The minimum current method recovers a good approximation 
to the original current density values in the boundary from the noisy 
data, except at locations affected by the noise. The method alters the 
boundary values for current density within the uncertainties. 
The minimum current method boundary values are shown also to
enable NLFFF reconstructions. The 
P and N solutions obtained from the boundary values are twisted 
bipole field configurations similar to the original NLFFF, but with 
reduced energies, due to the reduction in the boundary 
current density at locations affected by the noise. The self-consistency
procedure \citep{2009ApJ...700L..88W} is applied, and demonstrated to
produce a field which is a close match to the original NLFFF. The energy
of the self consistent solution is within a few percent of the energy of
the original NLFFF. The self-consistency procedure uses information 
on current density over both polarities of the boundary field, and hence
is able to recover current density values affected by noise at one or
other polarity.

We have chosen to minimize $J_z^2$ across the vector magnetogram to
produce boundary data suitable for NLFFF modeling. The vertical current 
density is a boundary condition for the force-free model, and large 
estimates of $J_z$ from vector magnetogram data (which may be due to 
observational errors) can prevent NLFFF codes from producing an 
accurate force-free solution. However, this choice may remove or 
reduce local current concentrations which are real, and which contribute
significantly to the energy of the field. Methods of
resolving the 180-degree ambiguity used to produce vector magnetograms
may also exclude real currents in the data (e.g.\ 
\citealp{1994SoPh..155..235M,2006SoPh..237..267M,2009SoPh..260...83L}). 
An advantage of 
the present method is that, if the 
uncertainties associated with currents are small, the currents are 
preserved. The accuracy of the minimum current estimate depends on the
accuracy of the uncertainties provided.

The test of the method presented here is highly idealised. It is not 
intended to provide a realistic example of the application of the 
method to vector magnetogram data, but only to demonstrate the method, 
and to show its performance in the best case. 

The method of inclusion of noise in the boundary data (adding Gaussian 
noise of known amplitude at specific locations to the horizontal field 
components of a NLFFF calculated from analytic boundary conditions) 
does not produce a close approximation to real solar data. We 
have not (yet) confronted the challenge of estimating electric currents 
from solar data with the minimum current method. Other studies have 
examined the influence of noise in vector magnetogram data \citep{2010A&A...511A...4W,2012SoPh..277...89L}. 
These studies add noise to synthetic line profiles, and then apply 
the various steps involved in constructing vector magnetograms 
(inversion, ambiguity resolution, co-ordinate transformations and 
rebinning, etc.). This process accurately mimics the influence of 
measurement errors on vector magnetograms. Tests of the method on
data of this kind may be tried in future.

The current density estimates presented here use centered differences, 
but higher order differencing schemes could in principle be used, 
following Equations~(\ref{eq:uijd})--(\ref{eq:sjz-general}).
High order schemes offer advantage in estimating gradients when 
applied to well-sampled data with small uncertainties, but the
advantage is lost in application to sparse data, or data with 
substantial errors, and these schemes have the disadvantage of 
introducing correlations in the neighbouring estimates. As such we
have chosen to use centered differences.

In future work the method will also be applied to vector 
magnetogram boundary data. This will allow investigation of whether the 
method is of use for coronal magnetic field modeling.

\begin{acks}[Acknowledgments]The author thanks the referee, Thomas 
Wiegelmann, for comments which helped improve this paper.\end{acks}

\end{article} 

\end{document}